\documentclass[journal]{IEEEtran}
\IEEEoverridecommandlockouts
\usepackage{cite}
\usepackage{amsmath,amssymb,amsfonts}
\usepackage{algorithmic}
\usepackage{graphicx}
\usepackage{textcomp}
\usepackage{xcolor}
\usepackage{caption}
\usepackage{subfigure}
\usepackage{booktabs} 
\usepackage{verbatim}
\usepackage{amsmath}
\usepackage{mathtools}
\usepackage{multirow}
\usepackage[linesnumbered,boxed,ruled,commentsnumbered]{algorithm2e}
\def\BibTeX{{\rm B\kern-.05em{\sc i\kern-.025em b}\kern-.08em
    T\kern-.1667em\lower.7ex\hbox{E}\kern-.125emX}}
\usepackage{authblk}
\usepackage{bbding}
\usepackage{algorithm2e}

\begin{document}

\title{Secure IoT Routing: Selective Forwarding Attacks and Trust-based Defenses in RPL Network\\
}

\author{Jun Jiang, Yuhong Liu\IEEEauthorrefmark{2}\thanks{\IEEEauthorrefmark{2}: Corresponding author}}
\affil{Department of Computer Science and Engineering, Santa Clara University, Santa Clara, CA, USA}
\affil{\{jjiang2, yhliu\}@scu.edu}

\maketitle

\begin{abstract}
IPv6 Routing Protocol for Low Power and Lossy Networks (RPL) is an essential routing protocol to enable communications for IoT networks with low power devices. RPL uses an objective function and routing constraints to find an optimized routing path for each node in the network. However, recent research has shown that topological attacks, such as selective forwarding attacks, pose great challenges to the secure routing of IoT networks. Many conventional secure routing solutions, on the other hand, are computationally heavy to be directly applied in resource-constrained IoT networks. There is an urgent need to develop lightweight secure routing solutions for IoT networks. In this paper, we first design and implement a series of advanced selective forwarding attacks from the attack perspective, which can flexibly select the type and percentage of forwarding packets in an energy efficient way, and even bad-mouth other innocent nodes in the network. Experiment results show that the proposed attacks can maximize the attack consequences (i.e. number of dropped packets) while maintaining undetected. Moreover, we propose a lightweight trust-based defense solution to detect and eliminate malicious selective forwarding nodes from the network. The results show that the proposed defense solution can achieve high detection accuracy with very limited extra energy usage (i.e. 3.4\%).
\end{abstract}

\vspace{2mm}

\section{Introduction}
With the rapid adoption of Internet of Things (IoT) devices around the world \cite{rose2015internet}, many of these devices are resource-constrained \cite{pereira2020challenges}. Since the existing Internet Protocols (IP) are too complex to be directly implemented on resource-constrained IoT devices \cite{marques2017energy}, the Internet Engineering Task Force (IETF) designs a lightweight IPv6 protocol with a series of core protocols to ensure efficient and secure communications, such as IPv6 over Low-power Wireless Personal Area Networks (6LoWPAN) \cite{le20126lowpan} and Routing Protocol for Low Power and Lossy Networks (RPL) \cite{winter2012rpl}. In particular, RPL, as the core routing protocol for resource constrained IoT networks, has been adopted by a variety of applications, such as healthcare \cite{gara2015rpl}, smart grid \cite{nassar2018multiple}, and smart city \cite{junior2020dynasti}, etc. 

Due to its popularity, RPL becomes an attractive attack target \cite{pongle2015survey}. One attack that can cause massive damage to the RPL network is selective forwarding attack\cite{wallgren2013routing}, where attackers interrupt network data flows by selectively dropping network packets. Compared to blackhole attacks which simply drop all packets, selective forwarding attacks are more deceptive and can remain undetected for a longer time, causing long-term damage to the network. Despite their impact, existing selective forwarding attacks are still \textit{lack of flexibility} in terms of dynamically identifying victim nodes and adjusting packet forward rates according to the state of the network.

On the other hand, there are three major types of defense mechanisms against selective forwarding attacks on RPL. The first type is to build a multi-path routing network to ensure the integrity of information transmission\cite{ma2017security}. These mechanisms often \textit{require excessive resources} to maintain the backup paths for nodes. The second type is distributed defense mechanisms, which deploy a monitor module on individual nodes\cite{airehrour2017trust}. However, these defense mechanisms often lead to \textit{significant extra energy consumption} at the monitoring nodes. In addition, it is challenging to ensure that all the monitoring nodes are \textit{long-term reliable and honest} in reporting their neighbors' behaviors. The third type is the centralized defense mechanisms \cite{raza2013svelte}, where a central node is employed to monitor and analyze malicious behaviors in the network. Nevertheless, the central node has to be deployed at a \textit{core location in the network} to ensure coverage across the entire network, and may be easily \textit{misled by complex bad-mouthing attacks}.

In this work, we aim to advance current studies from both the attack and defense perspectives. Specifically, from the attack aspect, we propose an advanced selective forwarding attack model, which can dynamically launch three types of malicious behaviors: (1) flexibly dropping packets from selected types of protocols, (2) adjusting the packet forward rate based on the average network packet forward rate to stay stealthy, and (3) dynamically selecting specific children nodes for bad-mouthing attacks. Furthermore, these attack behaviors can be combined to significantly increase the damages to the network and reduce the risk of being detected by state-of-the-art defense mechanisms. 

Furthermore, from the defense aspect, we propose a novel centralized trust-based defense mechanism to combat selective forwarding attacks in RPL networks. Compared to distributed defense mechanisms, the proposed scheme can significantly save the energy consumption for anomaly detection by deploying the defense computation only at the root node. Unlike other centralized defense schemes, the proposed scheme takes advantage of RPL's tree-based network topology to evaluate the trustworthiness of nodes and avoids the introduction of a trusted third-party node. Furthermore, by carefully designing the detection/notification module, the proposed scheme can effectively balance the trade-off between detection delay and energy consumption. 

The main contributions of this paper are as follows.
\begin{itemize}
\item This work proposes an advanced selective forwarding attack with three types of behaviors against RPL network. Malicious nodes in the proposed attack model cannot only flexibly choose the type of packets to drop, but also control the packet forward rates dynamically. As a result, these attackers are able to launch more stealthy attacks to avoid being detected. Furthermore, malicious nodes can also bad-mouth other normal nodes to cause false alarms in the system. The experiment results show that it can effectively evade RPL self-defense mechanism and state-of-the-art defense mechanisms.
\item This work proposes a novel centralized trust-based defense mechanism. In particular, the proposed trust model integrates a self-trust value, which reflects a node's trustworthiness in its packet forwarding behavior, and a tree-based descendant trust value, which takes advantage of the RPL network topology to prevent bad-mouthing attacks. Furthermore, the proposed defense mechanism is deployed on the root node, which can effectively reduce the total energy consumption caused by distributed anomaly monitoring scheme, and eliminate the security risks caused by the introduction of third-party devices. Experiment results show that the proposed scheme achieves high detection accuracy and low energy consumption.
\item This work proposes a novel anomaly report mechanism. Instead of using UDP packets, ICMPv6 control message is chosen to send information about malicious nodes. The reporting mechanism only starts when root node detects the malicious behaviors in the network. This reporting mechanism cannot only ensure that nodes in the network can be notified promptly, but also avoid causing information broadcast storms.
\end{itemize}
The rest of this paper is organized as follows. Section II discusses existing selective forwarding attacks and defense mechanisms in RPL networks. Section III introduces preliminaries of the RPL protocol. Section IV and V discuss the proposed selective forwarding attack and the defense mechanism in details. The results of the experiment are given in Section VI, followed by a conclusion in Section VII. 

\section{RELATED WORK}

\subsection{Selective Forwarding Attacks in RPL Networks}
RPL network faces a variety of security threats, which are mainly divided into three categories\cite{mayzaud2016taxonomy, almusaylim2020proposing}. The first type of attack is resource attack, such as flooding attacks\cite{le2013impacts} and increased rank attacks\cite{xie2010routing}. In these attacks, the attacker aims to exhaust the victim node's energy and reduce its lifetime by misleading it to execute a large number of unnecessary instructions. The second type of attack is traffic attack, such as sniffing attacks\cite{mayzaud2016taxonomy} and identity attacks\cite{wallgren2013routing}, where the attacker's main goal is to eavesdrop or manipulate the network's traffic. The third type of attack is network topology attack where the attacker undermines the security and stability of the network by changing the topology of the network, such as sinkhole attacks\cite{wallgren2013routing}, blackhole attacks\cite{raza2013svelte} and selective forwarding attack\cite{gara2017intrusion}.

Selective forwarding attack is one of most devastating type of attacks \cite{hu2014detection, mathur2016defence, ren2016adaptive}, which can cause severe damage to the network. However, in many of these attacks, the attackers set a fixed number of dropped packets or blindly attack all data packets, which increases the risk of being detected by defense mechanisms. In some advanced attacks\cite{wallgren2013routing, airehrour2017trust}, the proposed attacks interrupt the victim node's communications by only forwarding RPL control messages while dropping all \cite{wallgren2013routing} or partial data packets \cite{airehrour2017trust}. Different from these existing attacks, in this study, the proposed selective forwarding attacks can perform more flexible malicious behaviors and even bad-mouth other innocent nodes to mislead state-of-the-art defense schemes.  

\subsection{RPL Network Defenses}\label{sec:RPL_defense}
The original design of RPL protocol has some basic security schemes, such as the local and global repair mechanisms\cite{winter2012rpl}, which can be triggered by changes of network topology, e.g. a link failure. However, these basic security repair mechanisms are far from adequate to resist the rapid evolving security attacks\cite{airehrour2016securing}.

Beyond the basic repair mechanisms, there are mainly three categories of defenses. The first type is to establish multiple routing paths for each node to avoid selective forwarding attacks. In\cite{ma2017security}, the authors propose a secure routing protocol M-RPL, which establishes a hierarchical cluster network and backup paths for different clusters in the route discovery phase. In\cite{lodhi2015multiple}, the authors establish a temporary backup path for a node based on its packet delivery ratio. In\cite{jenschke2018multi}, the authors propose to use the principle of Packet Replication and Elimination (PRE), through IEEE 802.15.4 Time-Slotted Channel Hopping (TSCH) as media access, to create parallel paths from nodes to the root node. Although the multi-path mechanism can effectively resist attacks, due to the introduction of redundant network routes and extra resources to maintain the backup paths, these defense schemes often cause significant increases in nodes' energy consumption. 

The second type of defenses is distributed defense mechanisms that build monitor module on each network node. Due to its easy implementation, this type of defenses is the most popular one. In \cite{airehrour2017trust, pu2018mitigating, khan2017trust}, the authors propose various distributed trust-based mechanisms, where each individual node monitors its neighbors' incoming and forwarding traffic and calculates their trust values. In \cite{medjek2017trust, djedjig2020trust}, a distributed, collaborative and layered trust-based IDS (T-IDS) and a Metric-based RPL Trustworthiness Scheme (MRTS) are proposed respectively, where each node monitors and cooperates with its neighbors to detect and report intrusions. These distributed mechanisms, however, have to be deployed on each IoT nodes, leading to significant extra energy expenditure. On the other hand, it also presents a new challenge, as how resource-constrained nodes can provide long-term reliable and honest reports on their neighbors' behaviors.

The third type is centralized defense mechanisms. In\cite{raza2013svelte}, the authors use a SVELTE intrusion detection system (IDS), where a 6LoWPAN Mapper is placed on the IPv6 Border Router (6BR) to monitor and analyze the malicious behaviors in the network. In\cite{wallgren2013routing}, the authors propose a lightweight heartbeat protocol, in which the root node detects the malicious node by periodically exchanging an echo signal with all its children nodes. In \cite{ul2021ctrust}, authors introduce a control layer to achieve the hierarchical trust-based mechanism ``CTrust-RPL'' to monitor the nodes' behaviors. These centralized detection mechanisms are often energy-efficient since only a limited number of nodes are involved for behavior monitoring. However, since the root node cannot directly monitor each node in the network, it can be easily misled by some malicious attacks, such as bad-mouthing attacks \cite{gautam2020efficient}, where malicious nodes blame their parent/children nodes for packet dropping. As a result, it is challenging for such defense schemes to be robust against complex bad-mouthing strategies. 

In our prior conference paper\cite{jiang2018root}, we have studied blackhole attacks and proposed a centralized defense mechanism at the root node. On this basis, we further study the selective forwarding attacks and defense scheme in this work. Unlike the blackhole attack model, which blindly drops all packets, the attack model in this study launches attacks with three malicious behaviors, (1) flexibly select packet protocol, (2) dynamically adjust packet forward rate, (3) bad-mouth the victim children nodes. As a result, it can effectively hide the attack behaviors and cause long-term network damage. In addition, we propose a new lightweight centralized trust-based defense scheme to defend against selective forwarding attacks. Compared to our prior work, where root node simply uses the average packet forward rate as the trust value for each node, the defense scheme proposed in this work is more comprehensive by (1) integrating self-trust value and tree-based descendant trust value, (2) introducing a beta-based trust framework with a discounting factor to gradually reduce the impact of previous behaviors, and (3) assigning asymmetric discounting factors for good/bad behaviors, so that a node's bad behavior is remembered for a longer time. Furthermore, more comprehensive experiments are performed to evaluate the performance of the proposed attacks and defenses from different aspects.

\vspace{-2mm}
\section{Preliminaries: RPL Protocol}
As the attacks and defenses proposed in this work are based on RPL protocols, in this section, we briefly introduce some basis of RPL protocol. The Routing Protocol for LLN (RPL) is defined by the IETF's Routing Over Low power and Lossy Networks (ROLL) Working Group. In particular, each RPL network may contain multiple RPL instances. Each RPL instance may contain multiple Destination Oriented Directed Acyclic Graph (DODAG). In a DODAG, the root node, which is usually the most powerful node, is responsible for storing and managing the routing paths. Non-root nodes can be added to one or more DODAGs. 

RPL is a hierarchical-based routing protocol that relies on a DAG structure to exchange data among network nodes. Consequently, the parent and children relationship is essential for routing in a RPL network. Each DODAG has a specific objective function, which defines how each node selects its parent node. Based on the objective function, an optimal path from any leaf node to the root node can be constructed. 

\vspace{-2mm}
\subsection{Control Messages in RPL Network}\label{AA}
\begin{figure}[tb]
\centering
\includegraphics[width=3.45in]{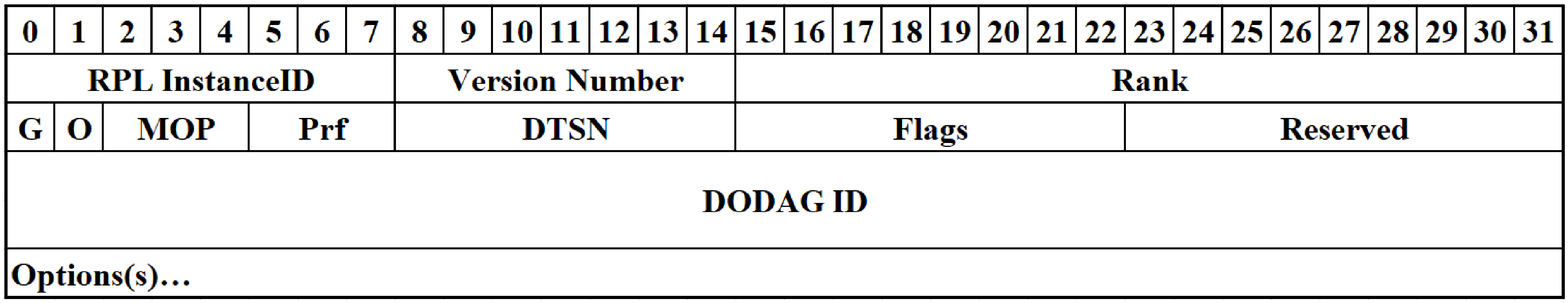}
\caption{Format of DIO message}
\label{fig:DIO_message}
\vspace{-4mm}
\end{figure}

There are four main types of control messages \cite{winter2012rpl} to establish the DODAG, which are DODAG Information Object (DIO), DODAG Information Solicitation (DIS), Destination Advertisement Object (DAO) and Destination Advertisement Object Acknowledgement (DAO-ACK). Particularly, DIO is the most frequently used message, sent by the root node to all other nodes in the network, to advertise network structures for DODAG discovery, assembly and maintenance. Therefore, to rapidly report anomaly detection results while avoiding extra overhead introduced by anomaly reporting messages, we propose to insert detection results into the DIO message to distribute to all nodes in the network. The format of a DIO message is shown in Figure \ref{fig:DIO_message}, which includes RPL InstanceID, Rank value, DODAG ID, Destination Advertisement Trigger Sequence Number (DTSN), etc. Non-root nodes must advertise and remain the values in DIO message, except for the update of the fields Rank and DTSN.

\subsection{Default Security Mechanisms in RPL Networks}\label{AA}
RPL networks can adopt standard mechanisms to ensure message integrity and confidentiality at different layers of the protocol stack. For example, the standardized IEEE 802.15.4 security, lightweight 6LoWPAN compression IPsec \cite{raza2011securing}, and Datagram TLS (DTLS) are adopted to ensure security at the data link layer, IP layer and transport layer, respectively. 

In addition, RPL adopts some simple repair mechanisms to recover from three network failures, including routing topology failure, link failure and node failure. When a small number of failures happen, the local repair mechanism starts. The local repair mechanism allows the nodes, which are impacted by failures, to detach from the original DODAG and change their rank values to infinite, then re-join the DODAG again. After multiple local repair mechanisms are initiated, the RPL protocol performs a global repair to rebuild the entire DODAG network by increasing the DODAG version number. Please note that the RPL uses a trickle timer to handle inconsistencies in the RPL DODAG. When the RPL network is stable, the interval of the trickle timer will increase exponentially. When a network inconsistency is detected, such as a loop generation, the trickle timer is reset. 

As briefly discussed in Section \ref{sec:RPL_defense}, these basic defense mechanisms are far from adequate when advanced attacks are launched.

\section{Proposed Selective Forwarding Attack}
Among diverse attacks against RPL networks, some attacks aim to cause as severe damage as possible to the network within a short time period, such as blackhole attacks\cite{jiang2018root}. These attacks are often easily detected and isolated by the defense mechanisms due to the aggressiveness of the malicious nodes' behaviors. Therefore, selective forwarding attacks, which can interrupt network communications in a flexible and stealthy way, are often launched to cause long term network damages. In this study, we propose an advanced selective forwarding attack model with three different types of selective behaviors.

\subsubsection{Protocol-based Attack}
We propose to selectively drop network packets according to their protocol types. Specifically, we propose to drop only data packets (i.e. non-ICMP packets) to achieve attack stealthiness. This is because in RPL networks, the loss of IPv6-based control messages (i.e. ICMP packets) will cause inconsistencies in network routing topology and trigger the RPL repair mechanisms. However, since data massages are transmitted based on UDP, the loss of such messages is difficult to be detected by RPL's self-defense mechanisms. In such cases, the RPL's self-recovery mechanisms will not be triggered \cite{wallgren2013routing}. 

As shown in Algorithm \ref{alg:PBS}, before the malicious node forwards a packet, it first determines whether the type of the packet is a data packet. If so, this packet can be dropped. Please note that, to enable selective behaviors based on protocol types, the malicious node only needs to check the header part of the messages, which will not incur significant processing power. 

\begin{algorithm}[tb]
    \caption{Protocol-based Attack}
    \label{alg:PBS}
    \LinesNumbered 
    \KwIn{$Packet$ (the Struct of IPv6 packet that need to be forwarded)}
    \If{$Packet.protocol \neq RPL$}
        {Drop packet\;}
\end{algorithm}

\subsubsection{Packet Forward Rate-based Attack}
We assume that the data messages transmitted between nodes are encrypted, so that the attacker can control whether the malicious nodes discard the data packet but cannot change the content of the data packet. The malicious nodes can dynamically adjust their packet forward rates (PFR) according to the network conditions. It is not easy to determine an appropriate PFR, which can cause non-trivial damages to the network while avoiding being detected. In this study, we propose to achieve this goal by controlling the PFR to be slightly above the average PFR of the network. Specifically, the malicious node estimates the average PFR by monitoring all its neighbors' incoming and outgoing packets, and ensures that its PFR is slightly above the average PFR of the network (i.e. by a small value $\varepsilon$). Please note this PFR will be dynamically updated based on the changes of the average network PFR. Consequently, the malicious node can hide itself while still causing long term damage to the network. The equation to calculate a node $N_i$'s PFR is shown below
\begin{equation}
\begin{aligned}
F_i^{\Delta t} = \frac{S_i^{\Delta t}}{R_i^{\Delta t}}
\end{aligned}
\label{eq:pfr}
\end{equation} 
where $R_i^{\Delta t}$ and $S_i^{\Delta t}$ represent the number of packets received and forwarded by node $N_i$ within time duration ${\Delta t}$, respectively. In the proposed attack, a malicious node estimates the network average PFR (i.e. $\bar{F}^{\Delta t}$) based on its neighbors' PFR, as shown below
\begin{equation}
\begin{aligned}
\bar{F}^{\Delta t} \approx \frac{\sum_{i=1}^p F_i^{\Delta t}}{p}
\end{aligned}
\label{eq:avg_pfr}
\end{equation}
where $p$ represents the number of neighbors of the malicious node. 

With a larger value of $p$ and a longer time duration $\Delta t$, the malicious node can achieve a more accurate estimation of $\bar{F}^{\Delta t}$, which, however, will also cause extra energy consumption and time delay. The proposed attack can flexibly adjust the trade-off according to specific attack scenarios. As shown in Algorithm \ref{alg:PFR}, before forwarding a packet, the malicious node determines whether to drop the packet by comparing the its current PFR with the observed network average PFR plus $\varepsilon$. The value of $\varepsilon$ can be adjusted according to the aggressiveness of the attack. 

\begin{algorithm}[tb]
\caption{Packet Forward Rate-based Attack}
\label{alg:PFR}
\LinesNumbered 
\KwIn{$Packet$ (the Struct of IPv6 packet that need to be forwarded), $F_m^{\Delta t}$ (PFR of the malicious node within duration $\Delta t$), $\bar{F}^{\Delta t}$ (Estimated network average PFR within duration $\Delta t$)}
\If{$F_m^{\Delta t}$ $ = $ ($\bar{F}^{\Delta t} + \varepsilon)$}{
Drop this packet\;}
\end{algorithm}

\subsubsection{Bad-mouthing Attack}
The proposed attack can arbitrarily choose one or multiple children nodes to achieve bad-mouthing attack. In bad-mouthing attack, the malicious node can frame the victim node (i.e. one of its children nodes) up by discarding data packets from the victim node. 

For this attack, the most challenging part is to selectively choose the victim nodes and attack strategy. Since blindly selecting children nodes to attack or attacking all children nodes increase the risk of the attacker being exposed to the detection mechanism, we propose to only select specific children nodes as the victim nodes. More importantly, the attacker can flexibly choose victim nodes that are either located at critical network positions, or requiring minimum attack effort. For example, as shown in Algorithm \ref{alg:bad_mouthing}, the malicious node identifies its children nodes with lower PFR as the victims as badmouthing these victims requires dropping less number of packets (i.e. less attack effort). 

After identifying the ideal victim node, the malicious node can dynamically discard the victim's packets, misleading the root node to identify the victim as a malicious node that drops packets. If multiple malicious nodes coordinately launch attack at the same time, the false alarm rate will significantly increase for most trust-based defense solutions. 

Please note that although we discuss these three attack behaviors independently for the sake of clarification, these attacks can be flexibly integrated to cause more damage. 

\begin{algorithm}[tb]
\caption{Bad-mouthing}
\label{alg:bad_mouthing}
\LinesNumbered 
\KwIn{$Packet$ (the Struct of IPv6 packet that need to be forwarded), $N_m$ (the Struct of malicious node), $\bar{F}^{\Delta t}$ (Estimated neighbor's average PFR), $ChList$ (list of children nodes), $NumVict$ (number of victim children nodes)}
sort $ChList$ in ascending order of PFR\; 
\For{$n=0;n \le NumVict; n++$}{
\If{$Packet.SrcMacAddr$ $ = $ $ChList[n].MacAddr$ and $Packet.DesMacAddr$ $ = $ $N_m.MacAddr$ and $F_{ChList[n]}$ $ >= $ $\bar{F}^{\Delta t}$}
{Drop packet\;
break\;}
}
\end{algorithm}

\vspace{-2mm}
\section{Lightweight Trust-based Defense Scheme}
\begin{figure}[tb]
\centering
\includegraphics[width=3.3in]{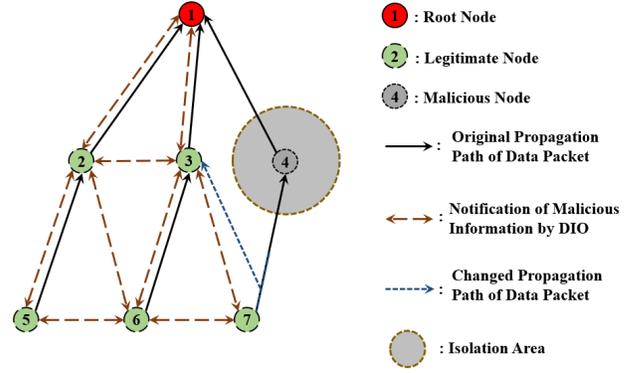}
\caption{Packet Propagation in the Proposed System}
\label{fig:system_overview}
\vspace{-4mm}
\end{figure}

In this section, we propose a lightweight trust-based defense scheme, which is deployed on the root-node, against selective forwarding attacks, as shown in Figure \ref{fig:system_overview}. The input for the defense scheme is the data packets received from non-root nodes. The defense scheme includes three major modules. The \textit{detection module} analyzes the trust value of each node based on the received data packets, whose propagation path is shown by the solid black line in Figure \ref{fig:system_overview}. After malicious nodes (e.g. node 4 in Figure \ref{fig:system_overview}) are identified, the \textit{notification module} encapsulates such information in DIO packets and notifies all the nodes in the network, as indicated by the brown dashed line in Figure \ref{fig:system_overview}. In the \textit{isolation module}, children nodes of the identified malicious nodes (e.g. node 7 in Figure \ref{fig:system_overview}) can isolate the malicious nodes and re-select their parent nodes based on received DIO messages. For example, the changed propagation path of data packets from node 7 is shown as the blue dashed line in Figure \ref{fig:system_overview}. 

In the rest of this section, we first introduce the design of the trust model, which is the core of the proposed defense solution. Then, we discuss each module of the proposed scheme in details.

\subsection{Trust Evaluation Model}
An advanced trust model is designed to evaluate the anomaly of each node's behavior. In a RPL network, the root node tracks the behaviors of each individual node $n_i$ and dynamically calculates the trust value, which is denoted as $T_i$. The trust value falls in the range from zero to one. When a node's trust value is below a trust threshold $\delta$, it will be identified as a malicious node. 

In the proposed trust model, the overall trust value $T_i$ of a node $n_i$ is composed of two parts: a self-trust value $T_{i} ^{s}$, which is to capture failures of packet forwarding, and a tree-based descendant trust value $T^{d}_{i}$, which is to capture bad-mouthing attacks.

\subsubsection{Self-trust Value}


The proposed scheme adopts the Beta trust model \cite{josang2002beta} as its basis. Beta distribution is a family of continuous probability distributions that are often used to model binary events. In our case, we consider whether a node can successfully forward a packet or not as a random binary event. Then based on the prior observations on the number of successful and failed packet forwarding events, the probability of this node to successfully forward the next packet can be estimated as the expected value of the beta distribution, as shown in equation (\ref{eq:beta_eq}), where $S$ and $L$ represent the number of packets successfully sent or lost by node $n_i$, respectively.
\begin{equation}
\begin{aligned}
T_{i}^{s} = \frac{S + 1}{(S + L + 2)}
\end{aligned}
\label{eq:beta_eq}
\end{equation}
From equation (\ref{eq:beta_eq}), we can observe that a node's self-trust value increases when more packets are successfully forwarded, or drops when more packets are lost.

However, the basic beta-based trust model cannot capture alternative behavior attacks \cite{labraoui2015off}, where malicious nodes alternatively perform good behaviors to accumulate high trust values and bad behaviors to interrupt network traffic. To prevent such attacks, we propose to introduce temporal information to discount a node's packet forwarding behaviors performed long time ago. Specifically, we introduce a discounting factor $f$ to gradually forget a node's behavior over time, so that a behavior with a smaller discounting factor indicates a lower influence on the node's trust value. The calculation of the discounting factor for the $k^{th}$ behavior of a node is shown below. 
\begin{equation}
\begin{aligned}
f_k = e^{-\lambda (q-k)}
\end{aligned}
\label{eq:kth_discount_factor}
\end{equation}
where $q$ represents the total number of behaviors performed by the node so far, including both successful and failed forwarding behaviors; and $(q-k)$ represents the total number of behaviors performed after behavior $k$. In other words, each time when a node performs a new behavior, all previous behaviors will have their $(q-k)$ value increased by 1, resulting in a smaller discounting factor. Please note that the latest behavior will always have its $\Delta_k = 0$, leading to its discounting factor value as 1. In addition, the forgetting speed $\lambda$ is a constant value. A larger $\lambda$ value will result in a smaller discounting factor and thus a higher forgetting speed. 

More importantly, to punish bad behaviors further, we propose to design an asymmetric trust model so that past good behaviors can be quickly forgotten while past bad behaviors will be remembered for a longer time. To achieve this goal, two different $\lambda$ values (i.e. $\lambda_g$ and $\lambda_b$) are adopted to separately discount past good and bad behaviors respectively, where $\lambda_g > \lambda_b$. Therefore, the self-trust value of node $i$ with $q$ behaviors (i.e. $T_{i,q}^s$) is calculated as follows.
\begin{equation}
\begin{aligned}
T_{i,q}^s = \frac{\sum_{k=1}^q {a_k f_{k,g}} +1}{\sum_{k=1}^q {(1-a_k) f_{k,b}}
+ \sum_{k=1}^q {a_k f_{k,g}} +2}
\end{aligned}
\label{eq:seif_trust}
\end{equation}
In equation (\ref{eq:seif_trust}), $a_k$ value is 1 if behavior $k$ is good, or 0 if behavior $k$ is bad.

\subsubsection{Tree-based Descendant Trust Value}
Due to the restricted hierarchical structure of RPL networks, only parent nodes will forward data packets for their children nodes. It is very easy for a malicious parent node to control the packet transmissions of one of its children nodes to launch bad-mouthing attacks against this child. By only considering self-trust value, the victim child node's trust will drop while the malicious parent's trust value remains the same. To further defeat such attacks, we propose to also introduce a tree-based descendant trust value for each node, which considers the trust value of its direct descendants. The descendant trust value $T^{d}_i$ is defined as follows.
\begin{equation}
\begin{aligned}
T^{d}_{i} = \frac{\sum(w_j T_{j}^s )}{C_i}
\end{aligned}
\label{eq:descendant_trust}
\end{equation}
where $T_{j}^s$ is the self-trust value of node $n_i$'s child node $n_j$. Parameter $w_j$ represents the weight of the child node $n_j$, which is determined by the number of data packet received by $n_j$ per time period. The greater number of data packets received by $n_j$ means larger weight assigned to node $n_j$. In addition, $C_i$ denotes the total number of packets received by node $n_i$. 

Please note that in the proposed scheme, a node's descendent trust value only depends on its children's self-trust value. Since RPL is a tree-like network, if a node's descendent trust value also considers its children's descendent trust values, it will lead to a recursive counting, where leaf nodes' trust values are over-emphasized. Since the descendant trust value is mainly designed to prevent bad-mouthing attacks, which can only effectively attack children nodes, we propose to not recursively count it. 

When the malicious node bad mouth any of its children nodes, its own descendant trust value will be decreased. Moreover, the descendant trust value will significantly decrease if the number of the victim children nodes increases. 

\subsubsection{Aggregated Trust Value}
Finally, the aggregated trust value $T_{i}$ of the node $n_i$ is the combination of the self-trust value $T_{i}^{s}$ and the descendant trust value $T^{d}_{i}$. The calculation of the aggregated trust value is shown below.
\begin{equation}
\begin{aligned}
T_{i} = w_s T_{i} ^{s} + w_d T^{d}_{i}
\end{aligned}
\label{eq:overall_trust}
\end{equation}
where $w_s$ and $w_d$ are the weights for self-trust value and descendant trust value. This aggregated trust value will serve as the major criteria to identify suspicious nodes in the network.

\subsection{Detection Module}
In this section, we present the detection module, which involves the above proposed trust model as its core. Specifically, we make two assumptions. First, the data messages transmitted between nodes are encrypted, meaning that malicious nodes on the routing path cannot change the content of the transmitted data information. Second, all data messages generated by non-root nodes are transmitted through the root node to the external network (e.g. the Internet). This is a reasonable assumption for most RPL networks \cite{wallgren2013routing, shreenivas2017intrusion}.

Current RPL protocol does not support the root node to record and track the packets sent by non-root nodes. To address this challenge, we introduce a sequence number, which is stored in the first byte of the data payload sent by each node. In particular, the sequence number is increased by one each time when the source node sends out a packet. The root node estimates the packet forward rate for each source node based on the number of received data packets and the corresponding sequence numbers. 

In addition, since frequently calculating the trust value for each node greatly increases the workload of the root node, we introduce a sliding time window. The root node only calculates the trust value of each node once in each sliding time window. The length of the window can be determined according to the network status, such as the battery capacity of the root node and the sensitivity of the trust value.

By calculating the trust value of each node, the root node can identify the possible malicious nodes according to Algorithm \ref{alg:MN_detection}. Confirmed malicious nodes will be added to the ``blacklist''. However, because the nodes may suffer from bad-mouthing attacks from their parent nodes, it may lead to high false alarm rate if we directly add all nodes with low trust values to the ``blacklist''. 

To reduce the false alarm rate caused by bad-mouthing attacks, we propose to add a ``watchlist'' and a trust recovery time period. When the trust value of a node is lower than the threshold for the first time, it will be added to the ``watchlist'' as a suspicious node. If a suspicious node is required to change its parent for further investigation, the root node will reset a recovery timer and track if the suspicious node's trust value can recover after the parent change action. Please note that the length of the recovery timer can be determined according to the specific network status. A longer timer leads to longer detection delay but lower false alarm rate. Within the recovery time period, if the trust value of the suspicious node recovers back to the threshold, indicating that changing its parent node stops the anomaly, this node is considered as a normal node. Then its parent will be identified as the malicious node and added to the ``blacklist''. Otherwise, the suspicious node is identified as a malicious node and moved to the ``blacklist''.

\begin{algorithm}[tb]
\caption{Malicious nodes detection}\label{alg:MN_detection}
\LinesNumbered 
\KwIn{trust threshold $T_{th}$, $ParentChangedList$ (The nodes in the list have changed parent.)}
\KwOut{$blacklist$ (containing malicious node ID),
$watchlist$ (containing suspicious node ID),}
INITIAL: $watchlist[]$ = empty, $blacklist[]$ = empty, $time period$ = constant value, $TrustRecoveryTime[]$ = constant value\;
\While{$time period$ expires}{
\ForEach{$j \in nodes$}{
\eIf{$T_{j}$ $ < $ $T_{th}$}{
\eIf{node $n_j$ not in $watchlist$}{
$watchlist$.add($n_j$)\;}{
\If{$ParentChangedList[j]$ = True and $TrustRecoveryTime[j]$ expires}{
$watchlist$.remove.($n_j$)\;
$blacklist$.add($n_j$);
$ParentChangedList[j]$ = False\;}} }{
\If{$ParentChangedList[j]$ = True and node $n_j$ in $watchlist$}{
$blacklist$.add($n_j$.old\_parent)\;
$watchlist$.remove($n_j$)\;
$ParentChangedList[j]$ = False\;}}}}
\end{algorithm}

\subsection{Notification Module}
After a malicious node is identified, the root node needs a reliable way to notify all the children nodes of the malicious node while avoiding information storms. This is challenging since the RPL network follows a strict tree-like topology for data packet forwarding, and a node can only receive data packets from its parent. It means that if data packets are used to disseminate the notifications, these packets will be simply dropped by malicious nodes and never reach their children nodes. 

\begin{figure}[tb]
\centering
\includegraphics[width=3.4in]{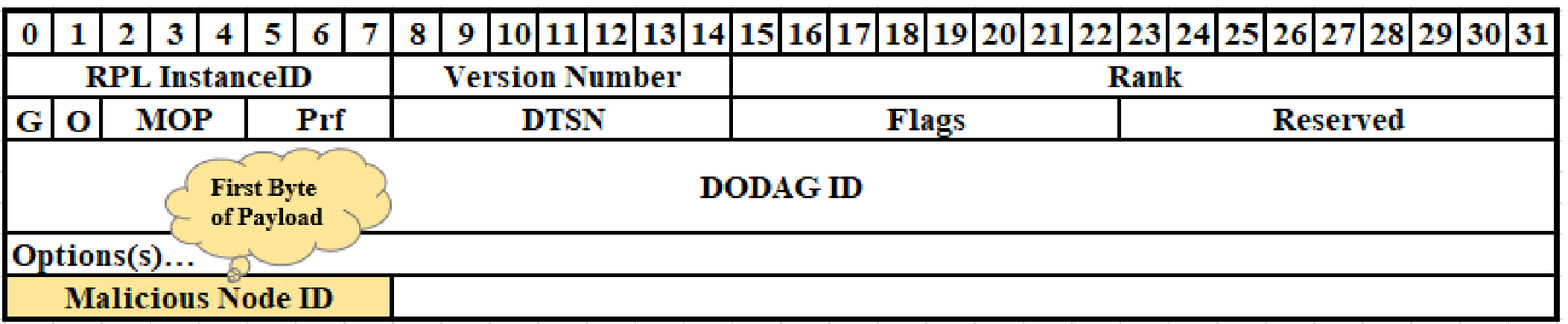}
\caption{The proposed notification message}
\vspace{-1em}
\label{fig:DIO_message_with_malicious}
\end{figure}

In this work, we propose to use the control messages (i.e. ICMPv6 messages) to disseminate these notifications, which can reach the children of a malicious node through other neighbor nodes. In particular, we recommend using the first byte in the payload of the ICMPv6 control messages to store the node ID, as shown in Figure \ref{fig:DIO_message_with_malicious}. The first bit can be used to distinguish the suspicious node, which is set to 0, and identified malicious node, which is set to 1. 

In order to reduce the energy consumption of the network, the root node executes the notification module right after the detection module. As shown in Algorithm \ref{alg:notification}, there are two lists: a ``blacklist'' and a ``ParentChangingList''. The root node marks the node information in the two lists separately. The ``blacklist'' is used to notify all the children nodes, whose parents are identified malicious nodes in the ``blacklist'', to re-select their parent node. The suspicious nodes, which are required to change parents, are added into the ``ParentChangingList''. This distinction can avoid unnecessary parent selection process for non-relevant nodes in the network.

In addition, the root node finds sub-trees, which can cover minimum number of nodes in the ``watchlist'', based on RPL tree-like topology. Then, the root node adds the nodes with the highest rank value in each sub-tree to ``ParentChangingList'', which is the function ``FindMaxRankNode'' in Algorithm \ref{alg:notification}.

\begin{algorithm}[tb]
\caption{Malicious nodes notification}\label{alg:notification}
\LinesNumbered 
\KwIn{$blacklist$, $watchlist$}
INITIAL:$ParentChangingList[]$ = empty, $ParentChangedList[]$ = False\;
\While{($watchlist$ not empty or $blacklist$ not empty) and $SendTime$}{
\If{$watchlist$ not empty}{
$ParentChangingList$.add\\(FindMaxRankNode($watchlist$))\;
broadcast($ParentChangingList$)\;
\ForEach{$i\in ParentChangingList$}{
$ParentChangedList[i]$ = True\;}}
\If{$blacklist$ not empty}{
broadcast($blacklist$)\;}}
\end{algorithm}

\subsection{Isolation Module}
As shown in Algorithm \ref{alg:Isolation}, when a non-root node in the network receives the notification message, it checks whether the node information in notification is its parent or itself. If any of two cases is true, this node removes the current parent from its parent list. Then, it re-selects its preferred parent and broadcasts this ICMPv6 control message to its neighbors. If not, the node broadcasts the ICMPv6 control message directly to all its neighbors. 

\begin{algorithm}[tb]
\caption{Isolation malicious nodes}\label{alg:Isolation}
\LinesNumbered 
\KwIn{$Packet$ (received ICMP6 control message which contains malicious node ID $M_{id}$ or suspicious node $SUS_{id}$), Parent ID $P_{id}$, Node ID $N_{id}$}
\eIf{$Packet$.$M_{id}$ = $P_{id}$ or $Packet$.$SUS_{id}$ = $N_{id}$}
{RplRemoveParent($P_{id}$)\;
PreferredParent = RplSelectParent($dag$)\;
broadcast($Packet$)\;}
{broadcast($Packet$)\;}
\end{algorithm}

After these three modules are completed, the malicious nodes launching selective forwarding attacks will be abandoned and isolated from the network.

\section{Experiment and Result Analysis}

\subsection{Experiment Set Up}
\begin{figure}[tb]
\centering
\includegraphics[width=3.1in]{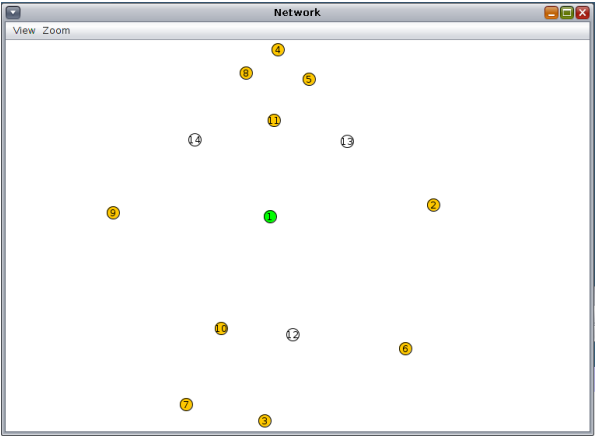}
\caption{Example of RPL Network Topology}
\label{fig:topology}
\vspace{-2mm}
\end{figure}

This work adopts Cooja, which is a network simulator of Contiki OS\cite{zikria2018survey}, as our experimental platform. Specifically, fifteen nodes are randomly deployed in the experiments as shown in Figure \ref{fig:topology}, including a root node (i.e. node 1), eleven legitimate non-root nodes (i.e. node 2 to node 11) and three malicious nodes (i.e. node 12 to node 14). Minimum Rank with Hysteresis Objective Function (MRHOF) is selected as the objective function, where children nodes select their preferred parents according to ranks and Expected Transmission Count (ETX) values. The simulation parameters are summarized in TABLE \ref{tab:parameters}. 

Based on the experiments, the performance of the proposed attack and defense scheme are tested and then compared with other state-of-the-art works. Specifically, this work adopts receiver operating characteristic (ROC) curve as the major performance metric because it can effectively reflect the trade-off between the detection rate and the false alarm rate when different thresholds are adopted. In each ROC curve, the x-axis and y-axis represent the detection rate and false alarm rate, respectively. The area under ROC curve (AUC) represents the accuracy. Larger area under the curve (i.e. the higher the AUC value) indicates better performance. 

\subsection{Performance of Proposed Attacks}
In this experiment, malicious nodes may either selectively drop the victim node's data packets based on their perceived average network PFR, or launch bad-mouthing attacks against a specific child node. To illustrate the impact of different attacks, the ratio of bad-mouthing attacks to the total number of attacks in the network is divided into four cases, 25\%, 50\%, 75\% and 100\%. 

The performance of the proposed attacks is evaluated against three defense schemes. In the first two schemes, the root node identifies the malicious nodes by comparing the PFR of each non-root node with a threshold value. The nodes with lower PFR are identified as malicious. The difference is that the first scheme (i.e. avg scheme) uses a node's average PRF, while the second scheme (i.e. rec scheme) uses only a node's most recent PFR. In the third scheme (i.e. def scheme), since the default RPL security scheme allows a node to re-select its parent node when network failures (e.g. link failure and node failure) are detected, the root node identifies malicious nodes by checking whether a node is discarded by its children.

\begin{table}[tb]
\renewcommand\arraystretch{1.3}
\centering
\caption{Simulation Parameters}
\begin{tabular}{|c|c|}
\hline
\textbf{Parameter}           & \textbf{Value}        \\ \hline
Simulation platform         & Contiki/Cooja 3.0     \\ \hline
MAC                          & CSMA/CA               \\ \hline
Transport                    & UDP/IPv6              \\ \hline
Emulated nodes               & Z1 mote               \\ \hline
Simulation coverage area     & 130 m * 130 m         \\ \hline
Total number of nodes        & 15                    \\ \hline
Malicious nodes              & 3                     \\ \hline
TX range                     & 50 m                  \\ \hline
Interference                 & range 100 m           \\ \hline
Packet size                  & 46-byte               \\ \hline
Data packet period           & 60 seconds            \\ \hline
Routing protocol             & RPL                   \\ \hline
Network protocol             & IP based              \\ \hline
Simulation time              & 150 minutes           \\ \hline
Link failure model           & UDGM with distance    \\ \hline
\end{tabular}
\label{tab:parameters}
\end{table}

\begin{figure}[tb]
\centering
\subfigure[25\% bad mouthing attacks]{
\begin{minipage}[t]{0.47\linewidth}
\centering
\includegraphics[width=1.7in]{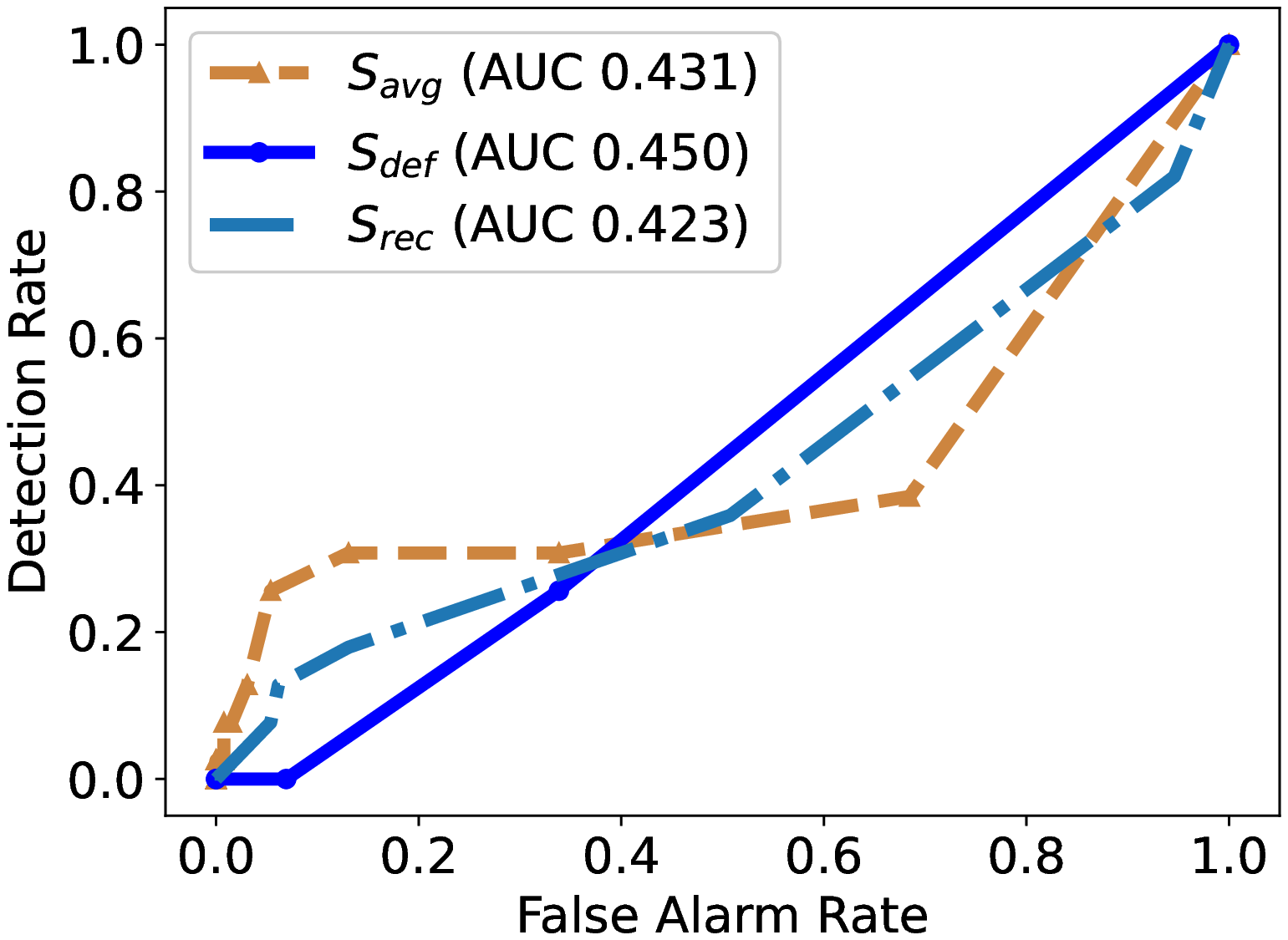}
\end{minipage}%
}%
\subfigure[50\% bad mouthing attacks]{
\begin{minipage}[t]{0.47\linewidth}
\centering
\includegraphics[width=1.7in]{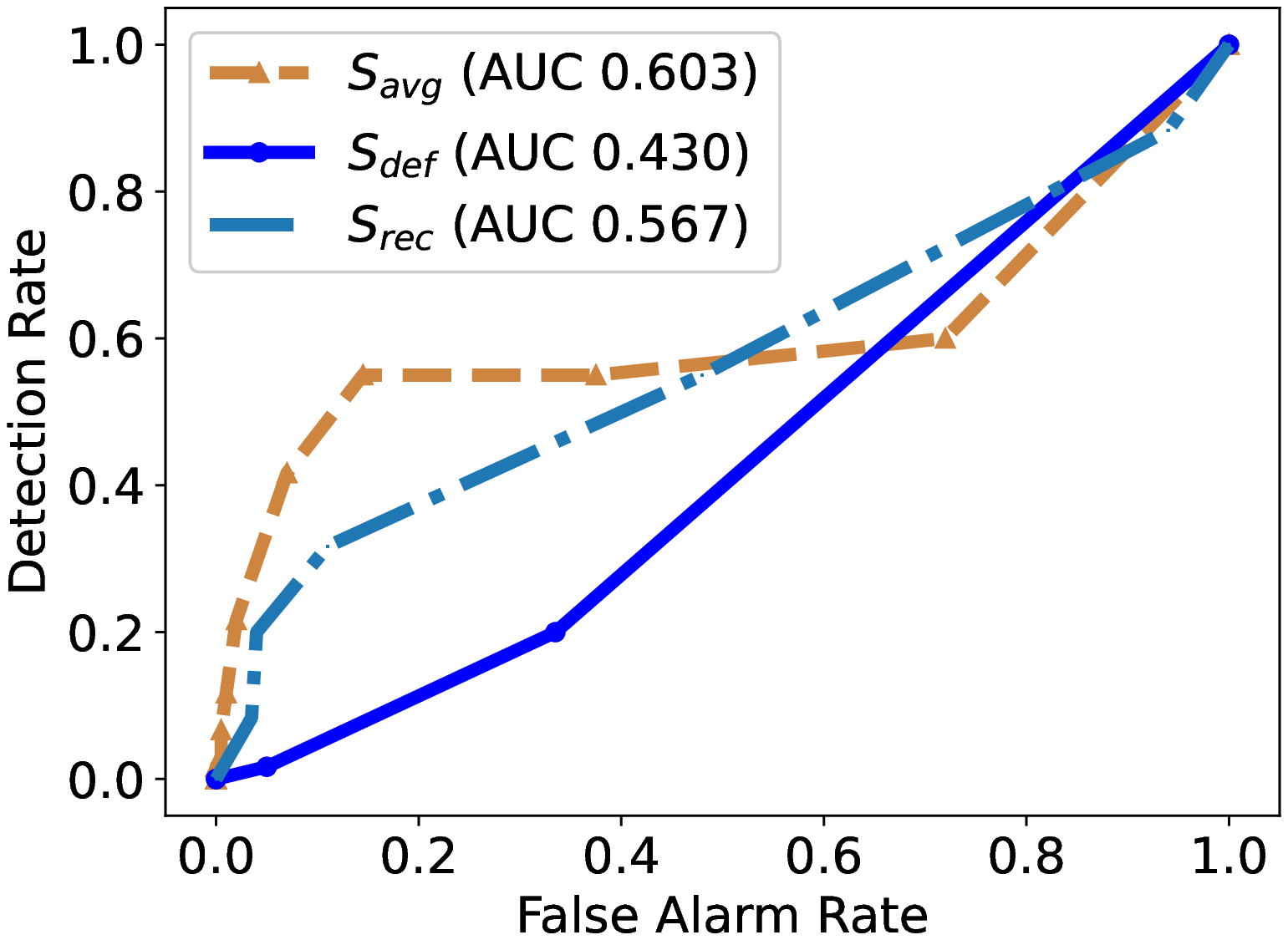}
\end{minipage}%
}%
\quad                
\subfigure[75\% bad mouthing attacks]{
\begin{minipage}[t]{0.47\linewidth}
\centering
\includegraphics[width=1.7in]{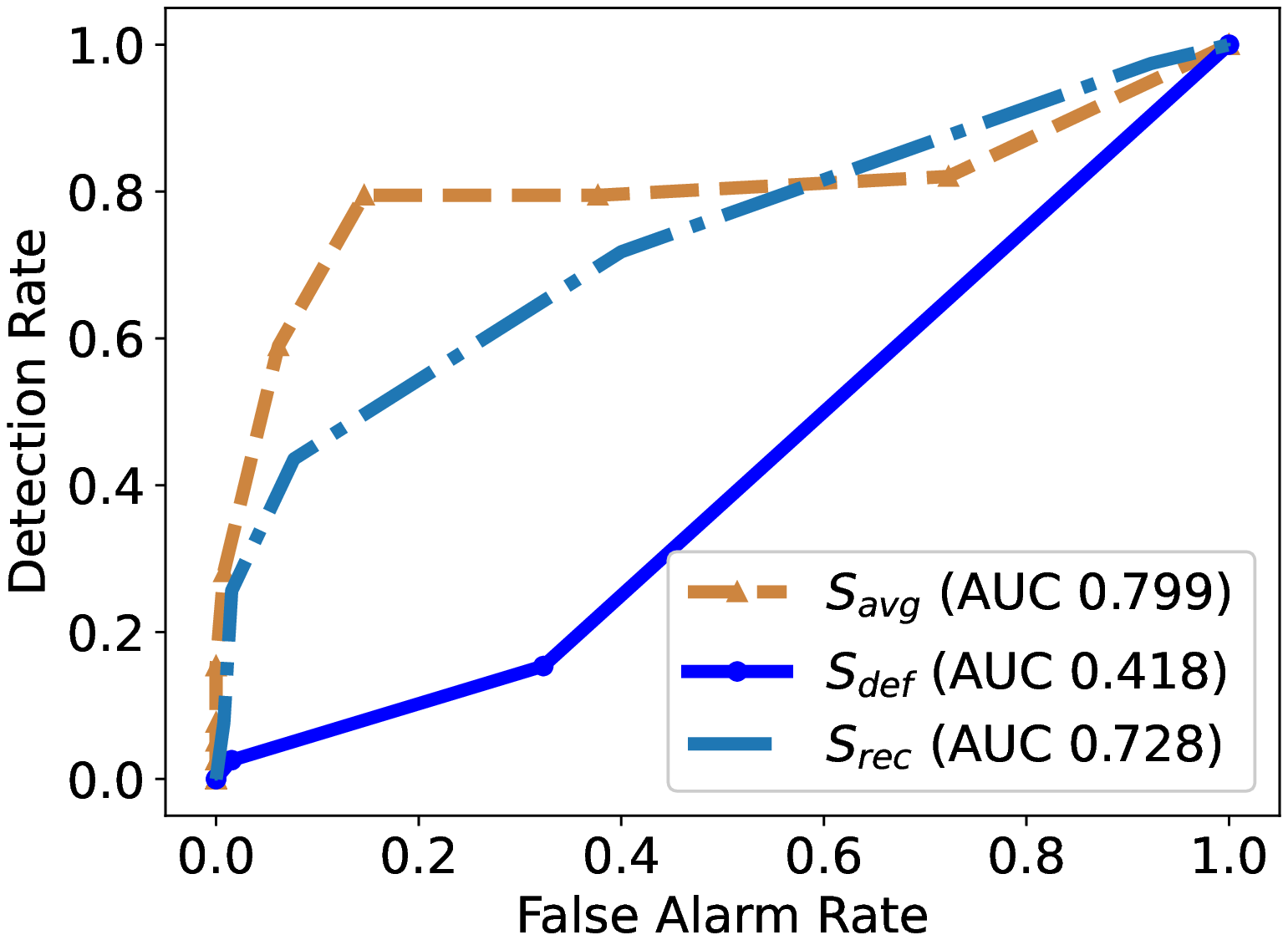}
\end{minipage}
}%
\subfigure[100\% bad mouthing attacks]{
\begin{minipage}[t]{0.47\linewidth}
\centering
\includegraphics[width=1.7in]{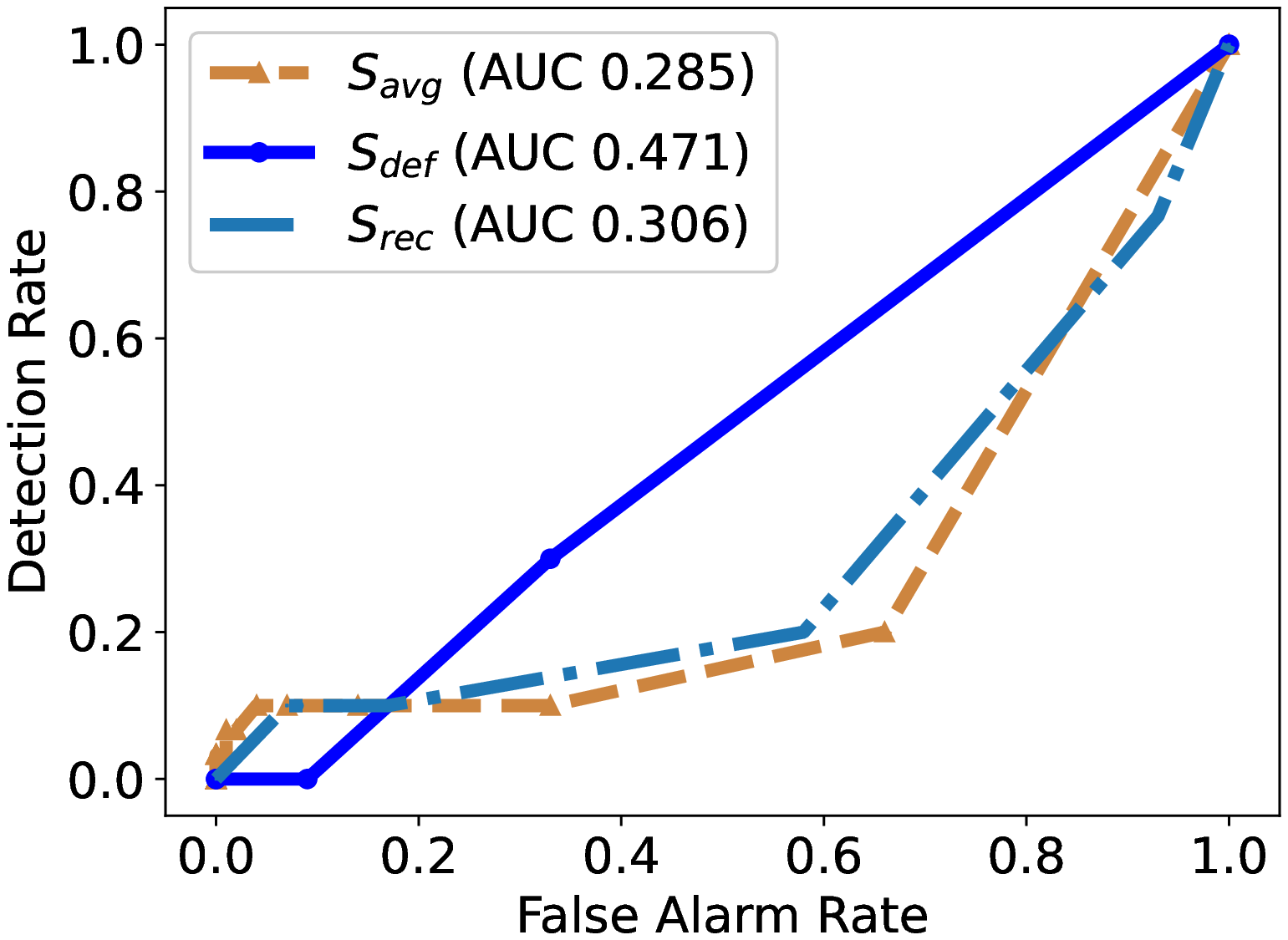}
\end{minipage}
}%
\centering
\caption{Performance of the proposed attack model against three defense schemes. $S_{avg}, S_{def}$ and $S_{rec}$ represents avg scheme, def scheme and rec scheme, respectively.}
\label{fig:attack performance}
\end{figure}

The effectiveness of proposed attack model is illustrated in Figure \ref{fig:attack performance}. In all sub-figures of Figure \ref{fig:attack performance}, the def scheme shows the lowest performance with 0.44 average AUC. This is because the proposed attacks only drop data packets, which rarely cause failures in network routing. Although avg scheme and rec scheme show slightly better performances (with an average AUC as 0.70 and 0.65 respectively in Figure \ref{fig:attack performance} (a)-(b)), their performances significantly drop (with an average AUC as 0.36) when the proportion of bad-mouthing attacks increases to above 75\%, as shown in Figure \ref{fig:attack performance} (c)-(d). This is because these two schemes consider a node with low PFR (i.e. either average PFR or the most recent PFR) as a malicious node, which can be taken advantage by bad-mouthing attacks to frame up the victim nodes.

\subsection{Performance of Proposed Defense Modules}
In this sub-section, we evaluate the effectiveness of each critical strategy proposed for the defense scheme. In particular, these strategies include (1) discounting factor, (2) asymmetric forgetting speed, and (3) integration of self-trust and descendant trust. Furthermore, the proposed attack models are launched with 50\% selective forwarding behaviors and 50\% bad-mouthing behaviors. 

\subsubsection{Effectiveness of Discounting Factor}
\begin{figure}[tb]
\centering
\includegraphics[width=3.5in]{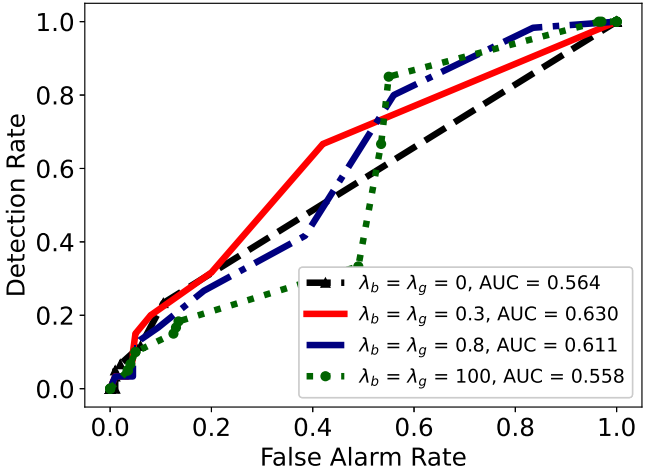}
\caption{Effectiveness of various forgetting speeds}
\vspace{-1em}
\label{fig:same_factor}
\end{figure}

In this subsection, $\lambda_g$ and $\lambda_b$ are set as the same value $\lambda$. By changing the values of $\lambda$, Figure \ref{fig:same_factor} illustrates its impact on the performance of the proposed defense scheme. Specifically, four different $\lambda$ values (i.e. 0, 0.3, 0.8, 100) are applied so that the discounting factor $f_k$ ranges in the interval $[0, 1]$. From Figure \ref{fig:same_factor}, it can be observed that the defense scheme shows the best performance (e.g. AUC = 0.630) when $\lambda = 0.3$.  

Specifically, when $\lambda = 100$, $f_k = e^{-\lambda (q-k)} = e^{-100(q-k)}$, which is approximately 0 for any $q-k>0$. It indicates that the defense scheme only remembers the most recent behavior for trust evaluation. In such settings, the defense scheme can mistakenly identify normal nodes with accidental packet losses as malicious nodes and therefore results in high false alarm rates, as shown by the green dotted curve in Figure \ref{fig:same_factor}. On the other hand, when the value of $\lambda$ is 0, the defense scheme, which remembers all previous behaviors, also performs worse, as shown by the blue dash dot line in Figure \ref{fig:same_factor}. This is because malicious nodes can easily mislead the defense scheme by accumulating high trust values through good behaviors performed long time ago. 

With an appropriate $\lambda$ value, the proposed defense scheme can achieve high performance. Particularly, as shown in Figure \ref{fig:attack performance} (b), when $\lambda = 0.3$, the AUC of the scheme with is 4\% and 11\% higher than that of the avg and rec schemes respectively, validating the effectiveness of the proposed defense scheme.

\subsubsection{Effectiveness of Asymmetric Forgetting Speeds} 
\begin{figure}[tb]
\centering
\subfigure[offset = 0.1]{
\begin{minipage}[t]{0.47\linewidth}
\centering
\includegraphics[width=1.7in]{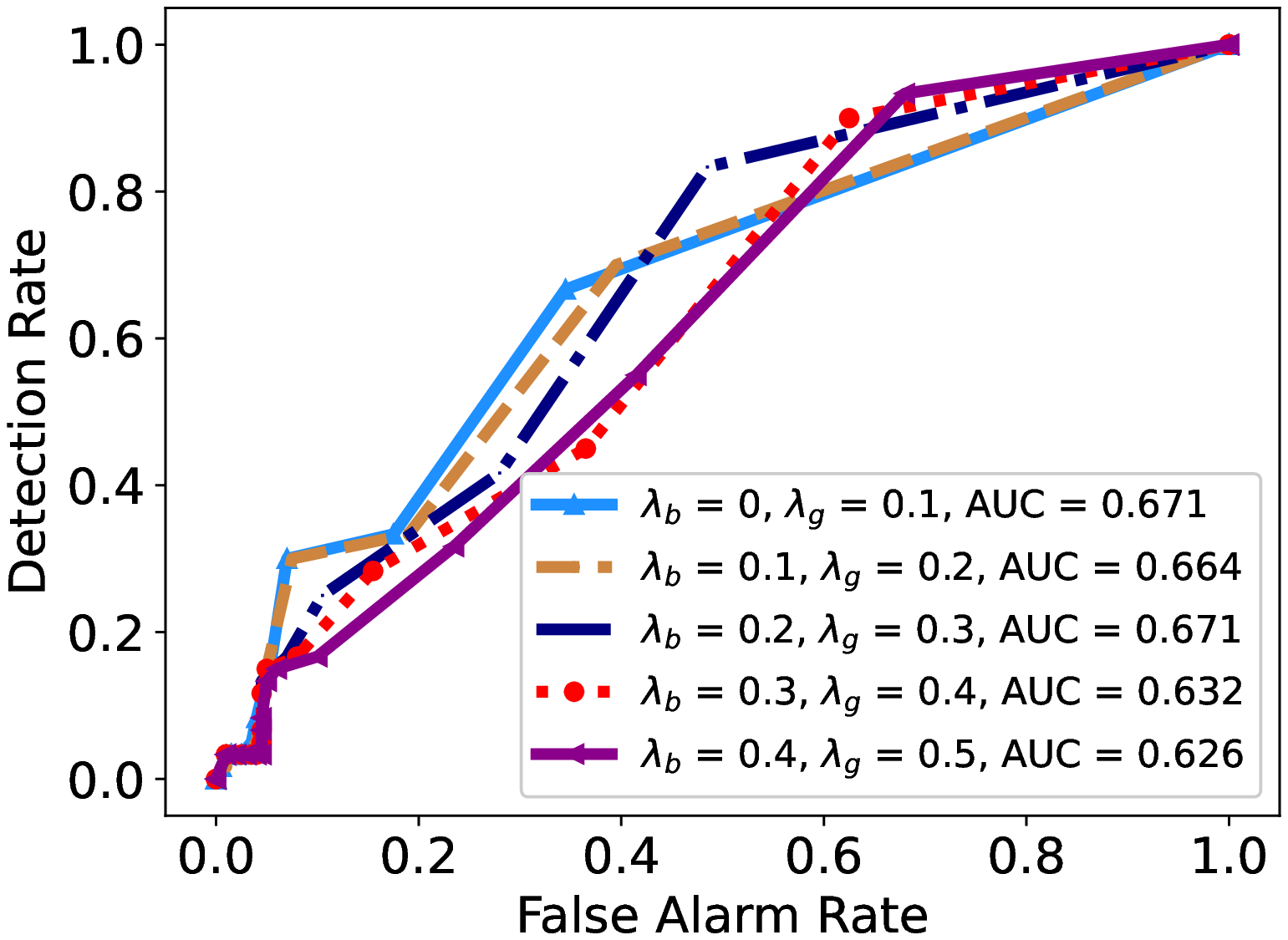}
\end{minipage}%
}%
\subfigure[offset = 0.2]{
\begin{minipage}[t]{0.47\linewidth}
\centering
\includegraphics[width=1.7in]{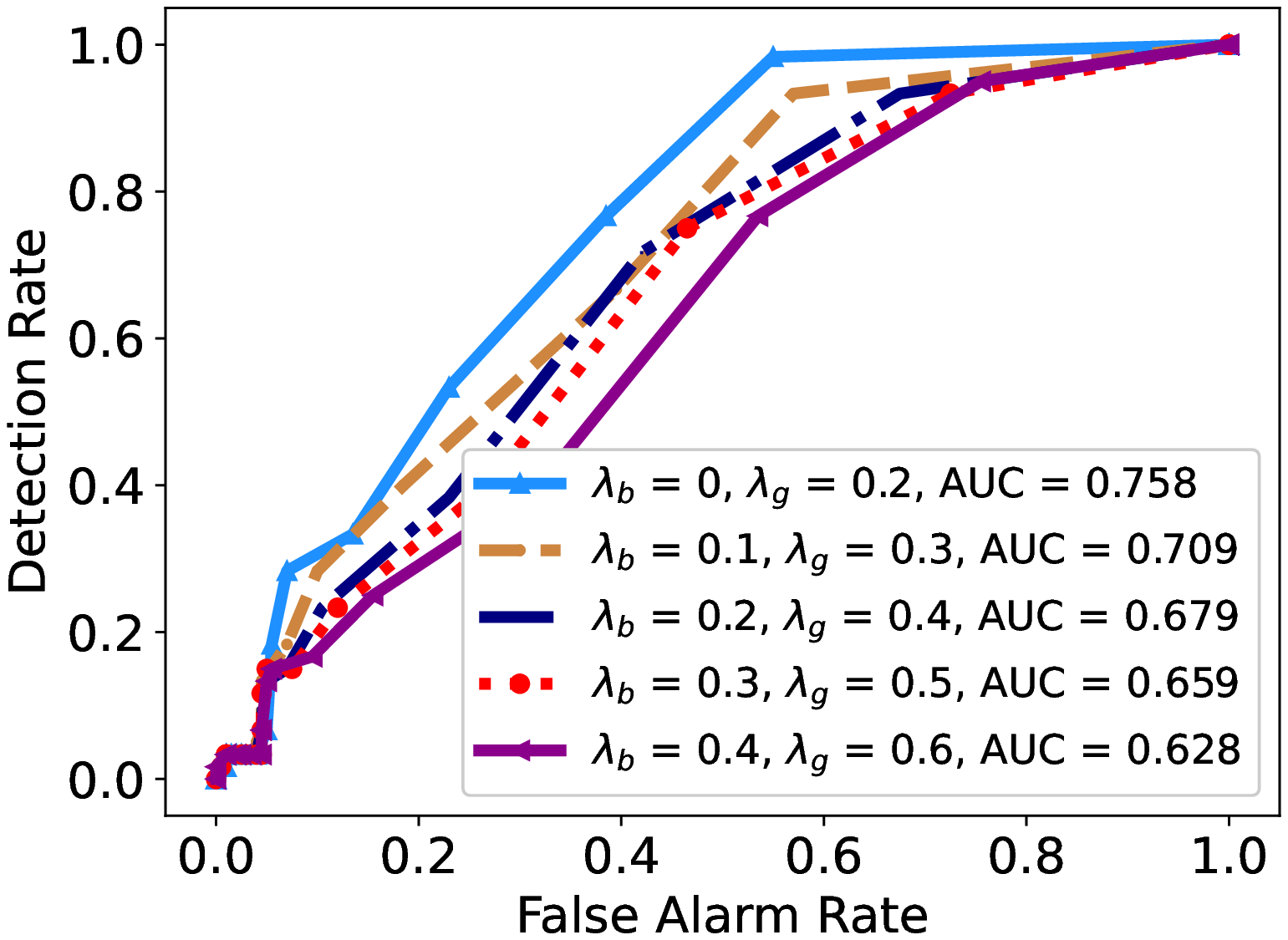}
\end{minipage}%
}%
\quad                
\subfigure[offset = 0.3]{
\begin{minipage}[t]{0.47\linewidth}
\centering
\includegraphics[width=1.7in]{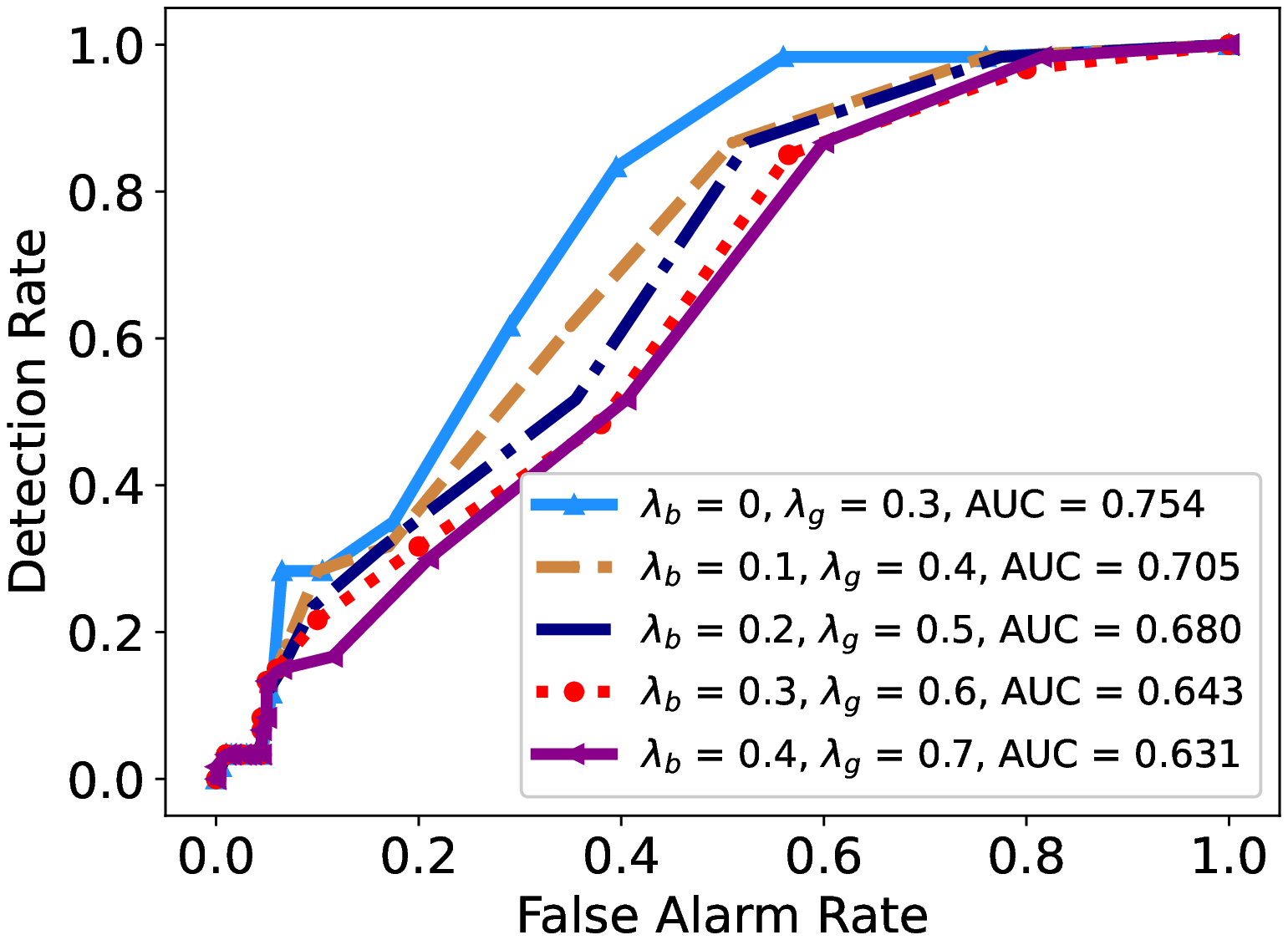}
\end{minipage}
}%
\subfigure[offset = 0.4]{
\begin{minipage}[t]{0.47\linewidth}
\centering
\includegraphics[width=1.7in]{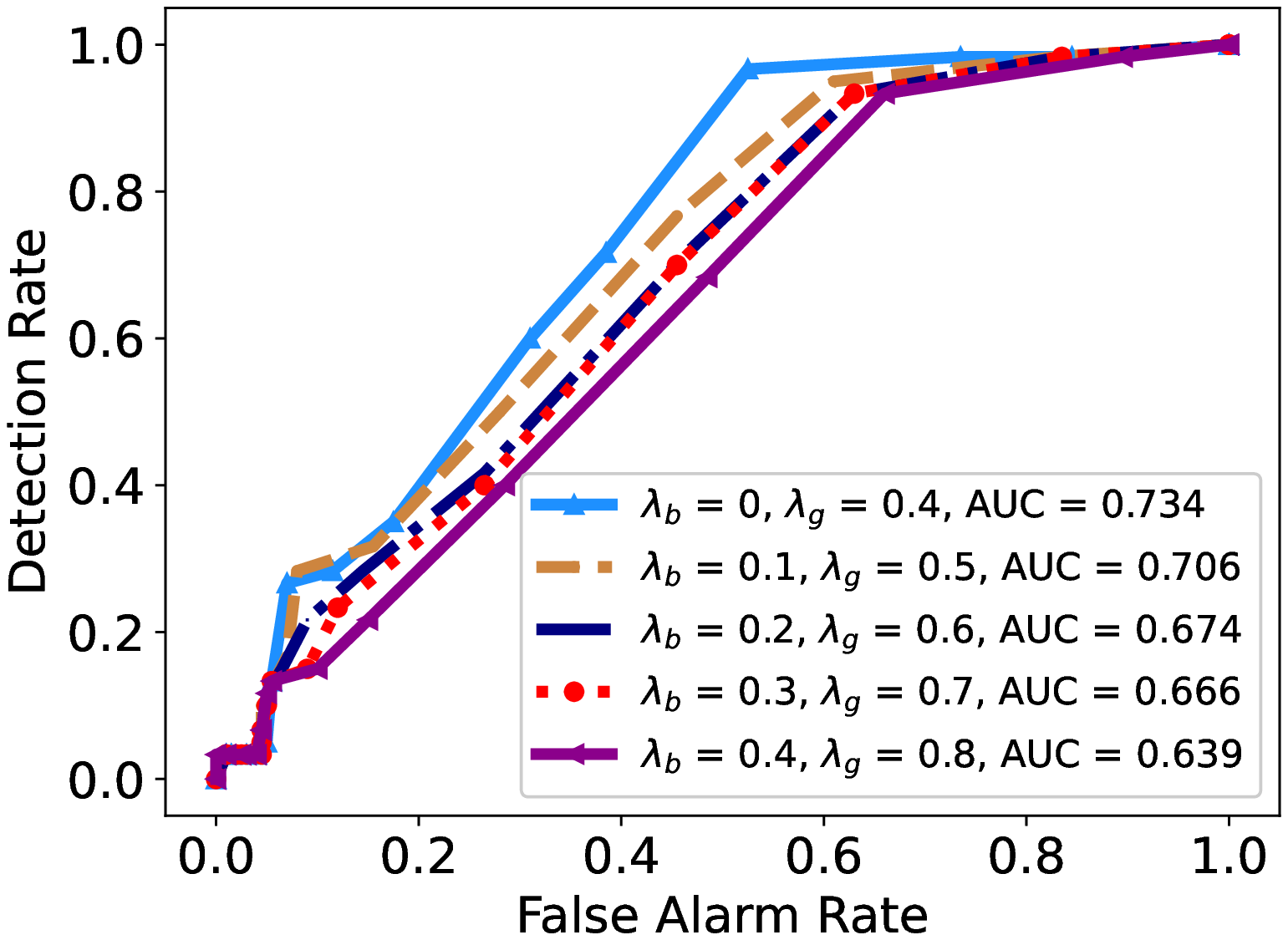}
\end{minipage}
}%
\centering
\caption{Effectiveness of asymmetric forgetting speeds. Offset represents the difference between forgetting speeds $\lambda_{b}$ and $\lambda_{g}$.}
\label{fig:different factors}
\end{figure}

Next, we evaluate the effectiveness of asymmetric forgetting speeds in Figure \ref{fig:different factors}. Observed from Figure \ref{fig:same_factor}, the defense scheme with $\lambda = 0.3$ shows the best performance. Therefore, in this subsection, the value range of $\lambda_{b}$ is set from 0 to 0.4. As we propose to forget bad behavior slower, the $\lambda_{g}$ values are set to be 0.1, 0.2, 0.3 and 0.4 higher than $\lambda_{b}$. The results are shown in Figure \ref{fig:different factors}. 

Comparing the schemes in Figure \ref{fig:same_factor} and Figure \ref{fig:different factors}, it can be observed that 79\% of the schemes with asymmetric forgetting speeds in Figure \ref{fig:different factors} perform better than the best scheme (i.e. $\lambda_{b} = \lambda_{g} = 0.3$) in Figure \ref{fig:same_factor}. In particular, the AUC of the best scheme (i.e. $\lambda_{b} = 0$ and $\lambda_{g} = 0.2$) in Figure \ref{fig:different factors} (b) is 20\% higher than that of the best scheme in Figure \ref{fig:same_factor}. This observation validates the effectiveness of the asymmetric forgetting speed.

Furthermore, an appropriate offset (e.g. 0.2) between the two forgetting speeds makes the scheme perform better. Specifically, the scheme with $\lambda_{b} = 0$ and $\lambda_{g} = 0.2$ in Figure \ref{fig:different factors} (b) performs the best with the AUC as 0.758, which is 13\%, 0.5\%, and 3\% higher than the highest AUCs in Figure \ref{fig:different factors} (a), (c), and (d). When the offset is too large, the scheme will be too sensitive to bad behaviors, which may increase the false alarm rate. When the offset is too small, the performance will be very similar to that of the scheme with identical $\lambda_b$ and $\lambda_g$, which is less effective.

\subsubsection{Effectiveness of Integrating Self-Trust and Descendant Trust}
To detect bad-mouthing attacks, where the parent node frames its children nodes by discarding their data packets, we propose to integrate self-trust ($T^s$) and descendant trust ($T^d$). This subsection aims to evaluate the trade-off by adjusting the weights of these two aspects. In particular, $w_s$ and $w_d$ represent the weights assigned to $T^s$ and $T^d$, respectively, where the sum of $w_s$ and $w_d$ equals one. The values of $w_s$ and $w_d$ are in the range of $[0.1, 0.9]$ with the step interval as 0.1. The values of $\lambda_{b}$ and $\lambda_{g}$ are fixed as 0 and 0.2 (i.e. the optimal values from Figure \ref{fig:different factors}), respectively.

\begin{figure}[tb]
\centering
\includegraphics[width=3.5in]{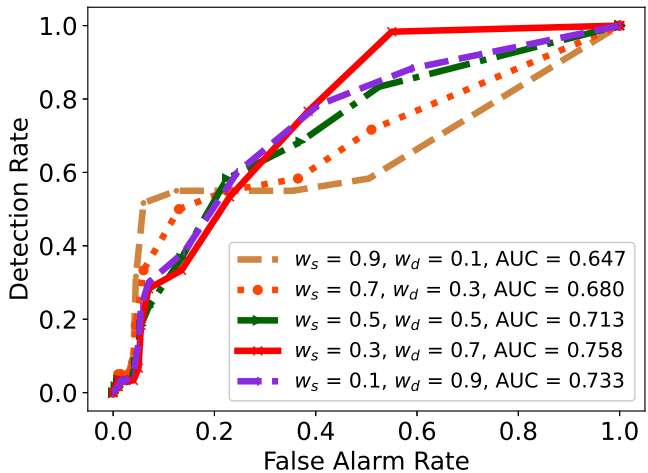}
\caption{Comparison of effectiveness of combination of two trust factors}
\vspace{-1em}
\label{fig:weight factor}
\end{figure}

From Figure \ref{fig:weight factor}, the proposed defense scheme with $w_s = 0.3$ and $w_d = 0.7$ achieves the best performance. Furthermore,  when $w_s \in [0.2, 0.4]$ and $w_d \in [0.6, 0.8]$, the AUC of the schemes are higher than 0.746, which validates the robustness of the proposed defense scheme. 

More importantly, the AUC of the best scheme in Figure \ref{fig:weight factor} is 26\% and 34\% higher than the avg scheme and rec scheme in Figure \ref{fig:attack performance} respectively. This observation validates that the introduction of descendant trust value ($T^d$) enables the proposed defense scheme to effectively detect bad-mouthing attacks in the network.

Some of the schemes in Figure \ref{fig:weight factor} have an inflection point in the detection rate range of $[0.55, 0.65]$, indicating that the increment of detection rate slows down. This is because this experiment places a limited number of nodes, including three malicious nodes, to prevent exceeding the capacity of the simulation platform (e.g. memory overflow). 


\subsection{Overall Performance Comparison}
This section compares the proposed defense scheme with the state-of-the-art defense schemes, including: (1) the RPL default recovering scheme: MRHOF \cite{winter2012rpl}, (2) a centralized scheme: Heartbeat protocol (HP) \cite{wallgren2013routing}, and (3) a distributed scheme: Trust-Aware RPL Routing Protocol (TPRP) \cite{airehrour2017trust}. All the schemes are applied on the same network topology and settings. In addition, the proposed defense scheme with $\lambda_{b} = 0$, $\lambda_{g} = 0.2$, $w_s = 0.3$ and $w_d = 0.7$ is launched.  

\subsubsection{Detection Accuracy}
\begin{figure}[tp]
\centering
\includegraphics[width=3.5in]{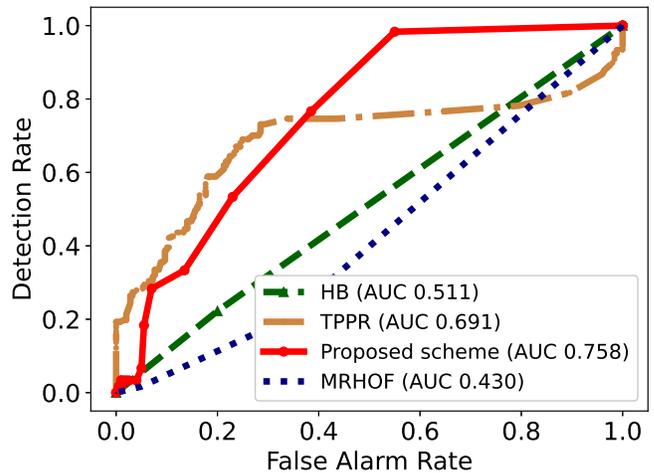}
\caption{Comparison of performance of different scheme based on proposed attack model}
\vspace{-1em}
\label{fig:overall_ROC}
\end{figure}

As shown in Figure \ref{fig:overall_ROC}, the proposed scheme yields the best overall performance among four defense schemes. Its AUC is 76\%, 48\% and 10\%, higher than that of MRHOF, HP and TPRP, respectively.

In particular, the HP and MRHOF schemes fail to differentiate malicious nodes from normal ones, because these two schemes detect malicious nodes based on replies of the control messages and thus cannot effectively capture the loss of data packets. Furthermore, when the false alarm rate is lower than 0.4, the detection rate of TPRP scheme stays high. This is because the TPRP scheme can effectively defend against bad-mouthing attacks by requiring each node in the network to monitor the sending and receiving packets of its neighbor nodes. However, the detection rate of TPRP scheme cannot be significantly improved. This is because it adopts average PFR as the trust value, which cannot effectively capture attacks alternatively perform good and bad behaviors.


\subsubsection{Energy Consumption and Detection Delay}
\begin{figure}[tb]
\centering
\includegraphics[width=3.5in]{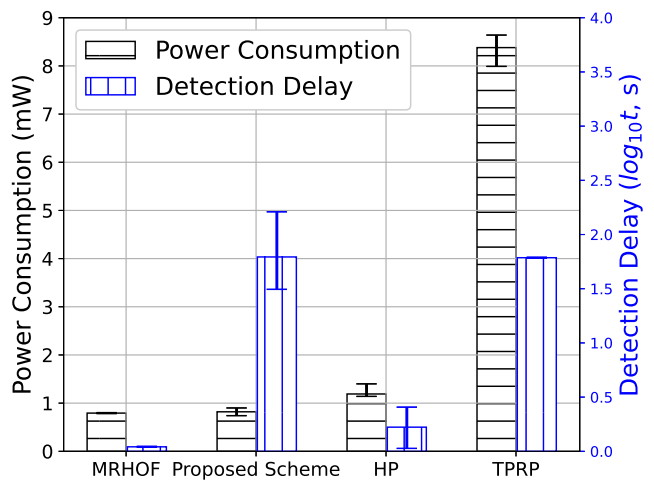}
\caption{Power consumption and detection delay comparison based on proposed attack mode}
\vspace{-1em}
\label{fig:power}
\end{figure}

Figure \ref{fig:power} compares the detection delay and energy consumption of legitimate non-root nodes for different defense schemes. Specifically, the proposed attacks are launched against these defense schemes. Each defense scheme has two bars, representing its energy consumption and detection delay respectively. 

From Figure \ref{fig:power}, we can observe that the proposed defense scheme consumes very limited power, only 3.4\% more than the default RPL recovery scheme (i.e. MRHOF). This is because the proposed scheme adopts the centralized design, where non-root nodes do not need to monitor neighbor nodes' activities, but only forward notification packets on detected malicious nodes. 

On the other hand, the HP and TPRP schemes are causing extra 50\% and 957\% power respectively when compared to the default MRHOF scheme. Although the HP scheme also adopts a centralized defense mechanism, it frequently launches a ``request and reply" process between the root node and non-root nodes, which increases power consumption. Furthermore, the power consumption of TPRP scheme is the highest due to its distributed design, which requires each non-root node in the network to monitor, analyze, and share the activities of its neighbors. These requirements significantly increase extra work time and computational costs for each node, leading to much higher power consumption. 

The detection delay is calculated based on only the successful detection of each scheme. In other words, if a detection scheme can only detect two malicious nodes out of three, the detection delay is the average delay of the two successful detection. As shown in Figure \ref{fig:power}, the MRHOF and HP schemes show small detection delay, which, however, is calculated based on the very limited malicious nodes that can be detected. Furthermore, there is a relatively large variation in the detection delay of the HP scheme. This is because the HP scheme relies on the exchange of ``request-reply'' messages to detect anomaly. The detection delay may vary based on the frequency of the request messages. A higher frequency may lead to smaller detection delay, but higher power consumption. 

In addition, the proposed scheme yields similar detection delay as the TPRP scheme, but a larger variation. This is because in the TPRP scheme, rather than relying on notifications from the root node, each node directly monitors its neighbors' behaviors, resulting in a relatively stable delay. The proposed scheme, however, can only detect anomaly when the data packets arriving at the root node show abnormal patterns, which may vary according to the network's data rate and the source node's network location. 

In summary, compared to the state-of-the-art defense schemes, the proposed scheme yields much higher detection accuracy (i.e. 10\% higher than the second highest one) and lower energy consumption (i.e. 31\% lower than that of the second lowest one). Although its detection delay is higher than other schemes (i.e. the MRHOF and HP schemes), it is practical to be applied in a low power low data rate RPL network. 

\section{Conclusion}
In RPL networks, malicious nodes can damage routing paths by selectively dropping packets. In this paper, we propose an advanced selective forwarding attack with three flexible attack behaviors, including protocol selection, packet forward rate selection and bad-mouthing with children nodes selection. The flexibility enables the malicious nodes to hide their attack behaviors and maximize the long term attack impact. Furthermore, we propose a new centralized trust-based defense scheme, which consists of a self-trust based on the beta trust model with asymmetric forgetting speeds, and a descendant trust value based on the RPL tree-like topology. Experimental results show that compared to the state-of-the-art defense solutions, the proposed defense scheme can effectively detect advanced proposed attacks with very limited energy consumption. 

\bibliographystyle{IEEEtran}
\bibliography{reference.bib}
\end{document}